\documentclass[superscriptaddress,twocolumn]{revtex4}
\usepackage{mathrsfs}
\usepackage{amssymb}
\usepackage[tbtags]{amsmath}
\usepackage{graphicx}
\usepackage{epsfig,graphicx,times}
\usepackage{color}
\usepackage{subfigure}

\begin{document}
\title{Weak values obtained from mass-energy equivalence}
\author{Miao Zhang\footnote{zhangmiao079021@163.com}}
\affiliation{Department of Physics, Southwest Jiaotong University,
Chengdu 610031, China}
\date{\today}

\begin{abstract}
Quantum weak measurement, measuring some observable quantities
within the selected subensemble of the entire quantum ensemble, can
produce many interesting results such as the superluminal phenomena.
An outcome of such a measurement is the weak value which has been
applied to amplify some weak signals of quantum interactions in lots
of previous references. Here, we apply the weak measurement to the
system of relativistic cold atoms. According to mass-energy
equivalence, the internal energy of an atom will contribute
its rest mass and consequently the external momentum of
center of mass. This implies a weak coupling between the internal
and external degrees of freedom of atoms moving in the free space.
After a duration of this coupling, a weak value can be obtained by
post-selecting an internal state of atoms. We show that, the weak
value can change the momentum uncertainty of atoms and consequently
help us to experimentally measure the weak effects arising from mass-energy equivalence.
\end{abstract}
\maketitle

\section{Introduction}
Almost 30 years ago, Aharonov, Albert, and Vaidman introduced
the theory of quantum weak measurement~\cite{AAV}. This measurement
has a key feature: after a weak coupling between the quantum
systems, the relevant observable quantities are measured in some
post-selected subensembles. An outcome of weak measurements is the
so-called weak value, $A_w$, which is a complex number and depends on
the post-selected state~\cite{Jozsa,imaginary,Anomalous}.
Physically, the pure state of a quantum system will be destroyed by
the postselection and then becomes a mixed state which is
constituted by two pure states. Selecting out one of the pure states
to be measured, some new phenomena appear because the selected wave
function is very different from the original one (before the
postselection). The postselection is actually a physical operation
applied on the quantum state, so that the weak measurement can be
regarded as a quantum coherent operation plus a classical selection
(filtering). Hence, the weak measurement is indeed different from
the usual measurement in physics.

Recently,
in refs.~\cite{Nature-wavefuction,Science-trajectory,HD,LG,Hall
Effect,CatNJPA,Cat,CatNJP,SuperluminalPTL, SuperluminalCherenkov,
SuperluminalBeryy,SuperluminalSteinberg,Electrons}, novel phenomena
within the weak measurements were reported, such as the spin Hall effect
of light~\cite{Hall Effect}, the Cheshire
cat~\cite{CatNJPA,Cat,CatNJP}, and the superluminal
phenomena~\cite{SuperluminalPTL, SuperluminalCherenkov,
SuperluminalBeryy,SuperluminalSteinberg}. Remarkably, lots of
studies have shown that the weak measurements can be utilized to
implement signal
amplifications~\cite{Amplification0,Amplification1,Amplification2,Amplification3,
Amplification4,Amplification5,Amplification6,AB,arXiv}. Most of
these references use a laser to realize the desired weak
measurements~\cite{RMP}. The necessary quantum interaction for weak
value gain was realized via the crystal-induced coupling between the
polarization and momentum of lights. Similar to the lights, the
matter wave should be also applicable for realizing the desired weak
measurements. Recently, this was demonstrated in neutron
interferometry~\cite{neutron}. In such an experiment, an external
magnetic field was applied to generate the spin-orbit coupling of
neutron, and the weak value was obtained by post-selecting a spin
state.

Here, we propose using weak value to test mass-energy equivalence
in the system of coherent atoms. Usually, the mass $m$ in the
Schr\"{o}dinger or Dirac equation is treated as a constant quality.
However, the mass-energy equivalence says that the internal energy of a particle will
contribute its rest mass~\cite{E=mc2}. Hence, the quality $m$ is no
longer a constant but depends on the particle's internal energies.
This implies a weak coupling between the internal and external
degrees of freedom of
particles~\cite{Zych1,Zych2,Pikovski,arXivDennis,Pollock}. The
potential coupling cannot be explained by the usual notation of
time-dilation~\cite{Hafele-Keating,Zhu2010,Mueller,1993} because the
rest-mass is now regarded as a quantum mechanics operator (for the
internal superposition states). A recent interesting
article~\cite{Pikovski} showed that the coupling due to the mass-energy equivalence is
perhaps an universal decoherence for quantum systems in the
gravitational field. As mentioned in Ref.~\cite{Zych1}, measuring
the possible dynamic coupling induced by mass-energy equivalence is still a challenge in
experiments due to its very weak interacting strength. Here, we
design a weak measurement setup for observing the predicted effects of mass-energy equivalence. We consider a two-level atom whose internal qubit and
external center-of-mass (c.m.) motion are coupled naturally due to
the mass-energy equivalence.
Based on this coupling, the weak value is obtained by
post-selecting the atomic internal state. We show that the present
weak value could offer some certain advantages for experimentally
detecting the ultraweak coupling induced by mass-energy equivalence.

This article is organized as follows. In Sec.~II, we briefly review
the basic principles and the usefulness of quantum weak
measurements. In Sec. III, we analyze the mass-energy-equivalence-induced coupling
between the internal and external degrees of freedom of a two-level
atom. Based on this coupling, the complex weak value is obtained by
post-selecting a proper internal state of atom. Considering the practical
atomic system is in an earth-based laboratory, the gravitational effect
is also studied. In Sec.~IV, we use the atomic Kapitza-Dirac (KD)
scattering to prepare the coherent atoms with large momentum
uncertainty~\cite{atomoptics}. Such a light-induced scattering of
atoms is further utilized to realize the desired postselection of
atomic internal states. Finally, we present our conclusions in Sec.
V.

\section{The basic principles for quantum weak measurements}
Consider two interacting quantum systems, namely Alice and Bob, with
the coupling Hamiltonian $\hat{H}=\hbar g_0\hat{A}\hat{B}$. Here,
$\hbar$ is the Planck constant divided by $2\pi$. $\hat{A}$ and
$\hat{B}$ are the operators that change respectively Alice and Bob's
states. The coupling strength between the two quantum systems is
$g_0$. After an interaction of duration $t$, the final state of the
total system can be approximately written as
\begin{equation}
|\psi\rangle=e^{\frac{-i\hat{H}t}{\hbar}}|A_i\rangle|B_i\rangle\approx|A_i\rangle|B_i\rangle
-ig\hat{A}\hat{B}|A_i\rangle|B_i\rangle\,
\end{equation}
under the weak interaction limit $g=g_0t\rightarrow0$. Here,
$|A_i\rangle$ and $|B_i\rangle$ are respectively the initial states
of Alice and Bob, and the high orders of $g$ have been neglected.
Considering Alice as a two-level system and then using its two
orthonormal states $|A_s\rangle$ and $|A_d\rangle$, the state (1)
can be expanded by
\begin{equation}
|\psi\rangle=(\langle A_s|A_i\rangle)|A_s\rangle|B_s\rangle+(\langle
A_d|A_i\rangle)|A_d\rangle|B_d\rangle
\end{equation}
with the unnormalized wave functions
$|B_s\rangle\approx(1-igA_w\hat{B})|B_i\rangle$ and
$|B_d\rangle\approx(1-igA^w\hat{B})|B_i\rangle$. Here, $A_w=\langle
A_s|\hat{A}|A_i\rangle/\langle A_s|A_i\rangle$ and $A^w=\langle
A_d|\hat{A}|A_i\rangle/\langle A_d|A_i\rangle$ are the weak values.
Numerically, when $\langle A_s|A_i\rangle\rightarrow0$, we have
$|A_w|\rightarrow\infty$ and $|A^w|\rightarrow\langle
A_i|\hat{A}|A_i\rangle$.

In weak measurement, the observable qualities of Bob are measured in
the post-selected subensemble of Alice's state $|A_s\rangle$, and the
relevant results are given by the following equation
\begin{equation}
\langle \hat{M}\rangle=\frac{\langle B_s|\hat{M}|B_s\rangle}{\langle
B_s| B_s\rangle}\approx\langle B_i|
\hat{M}|B_i\rangle+2g\text{Im}\left(A_wM_B\right)\,,
\end{equation}
with
\begin{equation}
M_B=\langle B_i|\hat{M}\hat{B}|B_i\rangle-\langle
B_i|\hat{M}|B_i\rangle\langle B_i|\hat{B}|B_i\rangle\,\,.
\end{equation}
Above, $\hat{M}$ is a Hermitian operator of Bob. It can be seen that
the $g$-induced effect is proportional to the imaginary part of
$A_wM_B$. This indicates that the weak value can amplify the
$g$-effect, significantly, with $|A_w|\rightarrow\infty$. It can be
also found that, the Eq.~(3) with $A_w=\langle
A_i|\hat{A}|A_i\rangle$ just describes the usual results without
the postselection. On quantum weak measurements, it is worth
emphasizing the following two points:

(i) The result $\langle \hat{M}\rangle$ is obtained in the
post-selected subensemble, it is not the conjoint-measurement
$\hat{P}=\hat{M}|A_s\rangle\langle A_s|$ performed in the entire
ensemble, i.e., $\langle
\hat{M}\rangle\neq\langle\psi|\hat{P}|\psi\rangle$.

(ii) The observable quality $\hat{M}$ can be generalized to be
$\hat{M}'=\hat{U}_2^\dagger\hat{M}\hat{U}_2$. That is, one can
perform any further operations, namely $\hat{U}_2$, on the selected
state $|B_s\rangle$ and then measure the observable quality
$\hat{M}$. The final result is $\langle B_s|
\hat{U}_2^\dagger\hat{M}\hat{U}_2|B_s\rangle/\langle B_s|
B_s\rangle$.

The only drawback of weak-value amplification is that the
probability $P_s\approx|\langle A_s|A_i\rangle|^2$ for successfully
post-selecting $|A_s\rangle$ decreased with the increasing of
$|A_w|$. Hence, the weak-value amplification is applicable just for
the quantum system which includes large numbers of microscopic
particles. Serval recent papers show weak-value amplification
offering no fundamental metrological advantage due to the
necessarily reduced probability of successful
postselection~\cite{SSNR,Fisher1,Fisher2}. Aiming at the practical
experimental systems, some studies show that the weak measurement
can still improve the precision of parameter estimation due to the
detector saturation~\cite{Jordan1} or some other systematic
errors~\cite{Zhu}. Alternatively, the advantage of weak-value
amplification can be simply explained by the following equation:
\begin{equation}
I=NP_s\langle \hat{M}\rangle+NP_sE_{\rm system}+E_{\rm random}\,\,.
\end{equation}
Here, $I$ is the experimental signal of observable quality $\langle
\hat{M}\rangle$. This signal is proportional to the number $NP_s$ of
successfully post-selected particles (with $N$ being the total
number of prepared particles). In practical experiments, the system
error $NP_sE_{\rm system}$ and the random error $E_{\rm random}$
will both contribute to the outcome $I$~\cite{Miao}. Certainly, the
random error $E_{\rm random}$ can be overcome by increasing the
number of particles, i.e., $NP_s\langle \hat{M}\rangle\gg E_{\rm
random}$. However, the system error cannot be eliminated by
increasing the particles input as it is proportional to $NP_s$.
Hence, the weak value amplification, i.e., $\langle
\hat{M}\rangle\gg E_{\rm system}$, could be useful for suppressing
the system errors (when the random errors are negligible).

As pointed out in the previous references, e.g.,Ref.~\cite{imaginary}, the
imaginary part $A_w^i$ of the weak value could be more useful than its
real part $A_w^r$. It can be also seen from Eq.~(3), when
$M_B=\text{Re}$, that the standard measurement (with $A_w=\langle
A_i|\hat{A}|A_i\rangle=\text{Re}$) does not work for measuring the
parameter $g$. This implies that the imaginary weak values could
reveal some new effects of $g$. In the following sections, we
show that the imaginary weak value can change the momentum
uncertainty of an atom moving in the free space.

\section{The mass-energy-equivalence-induced weak values}
\subsection{The dynamic evolution of a falling atom}
We consider only two internal states of an atom, namely, the ground
state $|g\rangle$ and the excited state $|e\rangle$. The ground and
excited states describe the different internal energies, i.e., $E_g$
and $E_e=E_g+\hbar\omega$, respectively. Here, $\omega$ is the
transition frequency between the two internal levels. According to
mass-energy equivalence, the atom has different rest mass for $|g\rangle$ and
$|e\rangle$, i.e., $m_g=E_g/c^2$ and $m_e=m_g+\hbar\omega/c^2$, with
$c$ being the speed of light in vacuum. Together with
the external c.m. motion, the total-energy of the atom can be expressed
as
\begin{equation}
\begin{aligned}
H&=\sqrt{c^2p^2+(m_g+m_c)^2 c^4}\\
&\approx(m_g+m_c)c^2+\frac{p^2}{2(m_g+m_c)}
\\
&\approx m_gc^2+\frac{p^2}{2m_g}+m_c(c^2-\frac{p^2}{2m_g^2})\,,
\end{aligned}
\end{equation}
with $m_c=0$ (for the ground state) or $m_c=\hbar\omega/c^2$ (for
the excited state). Here, $H$ is a classical Hamiltonian,
$p^2=p_x^2+p_y^2+p_z^2$ is the vector product of three dimensional
momentum, and the terms relating to the high orders of
$p^2/(m_gc+m_cc)^2$ and $m_c/m_g$ have been neglected for
simplification~\cite{Zych2}.

In quantum mechanics, the atom is described by a coherent
superposition of the two internal states, i.e., $|A_i\rangle=\alpha
|g\rangle+\beta |e\rangle$, with the probability-amplitudes $\alpha$
and $\beta$. Thus, the internal-state-dependent Hamiltonian should
be the following operator,
\begin{equation}
\hat{H}=(m_gc^2+\frac{\hat{p}^2}{2m_g})\hat{1}
+\hbar\omega[1-\frac{\hat{p}^2}{2(m_gc)^2}]|e\rangle\langle
e|\,,
\end{equation}
with $\hat{1}=|g\rangle\langle g|+|e\rangle\langle e|$ and
$\hat{p}^2=\hat{p}_x^2+\hat{p}_y^2+\hat{p}_z^2$. Therefore, the
results of measurements $\langle g|\hat{H}|g\rangle$ and $\langle
e|\hat{H}|e\rangle$ can satisfy the classical ones of Eq.~(6). By
neglecting the term $m_gc^2\hat{1}$ (which does not generate any
effects in the present work), the Hamiltonian (7) reduces to
\begin{equation}
\hat{H}=\hbar\omega|e\rangle\langle
e|+\left(1-\frac{\hbar\omega|e\rangle\langle
e|}{mc^2}\right)\frac{\hat{p}^2}{2m}\,.
\end{equation}
Here, we denote $m_g$ by $m$ for short (as it is frequently
used in what follows). Obviously, the coupling
$|e\rangle\langle e|\hat{p}^2$ induced by mass-energy equivalence depends basically on the parameter
\begin{equation}
g_0=\frac{\hbar\omega}{mc^2}\,,
\end{equation}
i.e., a ratio between the energy of a photon and the rest energy of
an atom (in the ground state). Such a ratio is very small, so the
relevant interactions are negligible in the usual
systems. For example, $g_0\approx8\times10^{-12}$ for the calcium
atom, with mass $m\approx6.7\times 10^{-26}$~Kg and the selected
optical transition frequency $\omega\approx4.6\times10^{14}$~Hz. In
the following, we suggest using the weak value to measure such an
ultrasmall parameter.

Note that the gravity effect of neutral atoms is significant in the
earth-based laboratory. Therefore, the above Hamiltonian should be
generalized to be
\begin{equation}
\hat{H}=\hbar\omega|e\rangle\langle e|+\left(1-g_0|e\rangle\langle
e|\right)\frac{\hat{p}^2}{2m}+\left(1+g_0|e\rangle\langle
e|\right)m\bar{g}\hat{z}\,.
\end{equation}
Here, the last term is due to the gravitational potential of the atom,
and $\bar{g}$ is the magnitude of gravitational acceleration on
earth. The coupling of $|e\rangle\langle e|\hat{z}$ is due to the
so-called Einstein's weak equivalence principle, i.e., the internal
energy $\hbar\omega$ will also contribute the gravitational
potentials to atom. In fact, the Hamiltonian (10) describes a
falling object with the different rest-masses which are internal-state dependent. The different rest masses (or say the different
objects) will lead to the different diffraction patterns in the
process of falling~\cite{1997,2004}. Some more detailed studies on
gravity-induced coupling can be seen in Refs.~\cite{Zych1,Zych2,Pikovski,arXivDennis}. In this
work, we do not measure the gravity effects. However, the relevant
studies on Hamiltonian (10) are necessary, because our suggested
measurements for coupling $|e\rangle\langle e|\hat{p}^2$ are
implemented in the earth-based laboratory.

In terms of $g_0$ couplings, we rewrite Hamiltonian (10) as
\begin{equation}
\hat{H}=\hat{H}_0-g_0\hat{H}_{rc}-g_0\hat{H}_{zc}\,,
\end{equation}
with a decoupled Hamiltonian
\begin{equation}
\hat{H}_0=\hbar\omega|e\rangle\langle e|+\frac{\hat{p}^2}{2m}
+m\bar{g}\hat{z}\,.
\end{equation}
This Hamiltonian is well known, describing a falling atom without
any couplings. The last two terms in Eq.~(11) describe the
$g_0$ couplings, by the horizontal directional Hamiltonian
\begin{equation}
\hat{H}_{rc}=\frac{\hat{p}^2_x+\hat{p}^2_y}{2m}|e\rangle\langle e|
\end{equation}
and the vertical Hamiltonian
\begin{equation}
\hat{H}_{zc}=\left(\frac{\hat{p}^2_z}{2m}-m\bar{g}\hat{z}\right)|e\rangle\langle
e|\,.
\end{equation}

Using Hamiltonian~(11), the Schr\"{o}dinger equation
$i\hbar\partial_t|\psi\rangle=\hat{H}|\psi\rangle$ can be rewritten
as $i\hbar\partial_t|\psi'\rangle=\hat{H}_{\rm int}|\psi'\rangle$
with an interacting Hamiltonian of
\begin{equation}
\hat{H}_{\rm int}=-g_0e^{\frac{i\hat{H}_0t}{\hbar}}
(\hat{H}_{rc}+\hat{H}_{zc}) e^{\frac{-i\hat{H}_0t}{\hbar}}\,.
\end{equation}
Above, $|\psi\rangle= \exp{(-i\hat{H}_0t/\hbar)}|\psi'\rangle$ is
the time-dependent state of atom, and $|\psi'\rangle$, the atom's
state in the interacting picture, will be resolved by the
Schr\"{o}dinger equation with Hamiltonian (15). We note that, the
unitary transformation $\hat{U}_0=\exp{(-i\hat{H}_0t/\hbar)}$ does
not generate coupling between the atomic internal and external
degrees of freedom, and therefore the key question is to solve the
state $|\psi'\rangle$. Indeed, to give an exactly resolved-result of
$|\psi'\rangle$ is rather difficult due to the presence of
gravitational potentials in $\hat{H}_0$ and $\hat{H}_{zc}$. However,
the question can be greatly simplified if one use the weak
measurement theory, i.e., considering only the effects from the
linear $g_0$.

According to the commutation relation
$[\hat{H}_{0},\,\hat{H}_{rc}]=0$, Eq.~(15) reduces to
\begin{equation}
\hat{H}_{\rm int}=
-g_0\hat{H}_{rc}-g_0e^{\frac{i\hat{H}_0t}{\hbar}}\hat{H}_{zc}
e^{\frac{-i\hat{H}_0t}{\hbar}}\,,
\end{equation}
and where the last term can be expanded by the well-known form of
\begin{equation}
\begin{aligned}
e^{\frac{i\hat{H}_0t}{\hbar}}\hat{H}_{zc}
e^{\frac{-i\hat{H}_0t}{\hbar}}&=\hat{H}_{zc}
+\frac{it}{\hbar}[\hat{H}_0,\hat{H}_{zc}]\\&\,\,\,\,\,\,+\frac{1}{2!}
\left(\frac{it}{\hbar}\right)^2[\hat{H}_0,[\hat{H}_{0},\hat{H}_{zc}]]
+\cdots\\
&
=\hat{H}_{zc}-\bar{g}t\left(2\hat{p}_z-m\bar{g}t\right)|e\rangle\langle
e|
\\
&=\hat{H}'_{zc}\,.
\end{aligned}
\end{equation}
Consequently, the interacting Hamiltonian (16) reads $\hat{H}_{\rm
int}=-g_0(\hat{H}_{rc}+\hat{H}'_{zc})$, with a time-dependent
Hamiltonian $\hat{H}_{zc}'$ in the $z$ direction.

Using the Dyson evolution operator
\begin{equation}
\begin{aligned}
\hat{U}=&1+\frac{-i}{\hbar}\int_0^t\hat{H}_{\rm int}dt_1\\
&+ \left(\frac{-i}{\hbar}\right)^2\int_0^t\hat{H}_{\rm
int}\int_0^{t_1}\hat{H}_{\rm int}dt_2dt_1+\cdots\,,
\end{aligned}
\end{equation}
we have the time-dependent state
\begin{equation}
\begin{aligned}
|\psi'\rangle&=\hat{U}|\psi_i\rangle \\
&=\left[1+\frac{ig_0}{\hbar}
\left(\hat{H}_{rc}t+\int_0^t\hat{H}'_{zc}dt_1\right)+o(g^2_0)\right]
|\psi_i\rangle\,.
\end{aligned}
\end{equation}
Here, the high orders of $\hat{H}_{\rm int}$ (i.e, the high orders
of $g_0$) have been neglected. $|\psi_i\rangle=|i\rangle|A_i\rangle$
is the initial state of the atom, with the internal $|A_i\rangle$ and
the external $|i\rangle=|\phi_x\rangle|\phi_y\rangle|\phi_z\rangle$.
Transforming back to the Schr\"{o}dinger picture, the state reads
\begin{equation}
|\psi\rangle=\hat{U}_0|\psi'\rangle
=\hat{U}_0\left[1+\frac{ig_0t}{\hbar}\hat{H}_{\rm
eff}|e\rangle\langle e|+o(g^2_0)\right]|\psi_i\rangle,
\end{equation}
with
\begin{equation}
\hat{H}_{\rm eff}=\frac{\hat{p}^2}{2m}-\hat{G}\,,
\end{equation}
and where
$\hat{G}=m\bar{g}\hat{z}+\hat{p}_z\bar{g}t-m(\bar{g}t)^2/3$. These
equations indicate that the kinetic energy and the gravitational
potential of atom will both contribute the entanglement between the
internal and external degrees of freedom. However, there is no
directly coupling (such as $\hat{p}_x^2\hat{z}$ and
$\hat{p}_x^2\hat{p}_z$) between the external c.m. motions for the
linear $g_0$. This greatly simplifies our calculation.

\subsection{The weak values due to the postselection of atomic
internal states}
Based on the entangled state (20), we post-select
an internal state $|A_s\rangle$; then the external motion collapses
on the state of an unnormalized form of
\begin{equation}
\begin{aligned}
|f\rangle&=\langle
A_s|\psi\rangle\\
&=\left[\langle A_s|\hat{U}_0|A_i\rangle+\frac{ig_0t}{\hbar}\langle
A_s|\hat{U}_0|e\rangle\langle e|A_i\rangle\hat{H}_{\rm
eff}+o(g^2_0)\right]
|i\rangle\\
&=\langle A_s|A'_i\rangle\hat{V}_t\left[1+\frac{iA_wg_0t}{\hbar}
\hat{H}_{\rm
eff}+o(g^2_0)\right] |i\rangle\\
&=\langle A_s|A'_i\rangle\hat{V}_t\left[1+\frac{iA_wg_0t}{2m\hbar}
(\hat{p}^2-2m\hat{G})+o(g^2_0)\right] |i\rangle\,.
\end{aligned}
\end{equation}
Here,
\begin{equation}
A_w =\frac{\langle A_s|e\rangle\langle e|A'_i\rangle}{\langle
A_s|A'_i\rangle}
\end{equation}
is our weak value, with $|A'_i\rangle=\exp(-i\omega
t|e\rangle\langle e|)|A_i\rangle$. The operator
$\hat{V}_t=\exp\{-i[\hat{p}^2/(2m)+m\bar{g}\hat{z}]t/\hbar\}$
describes the free evolution of atomic external motions in duration
$t$. The first result from the $g_0$ couplings can be found in the
postselection probability,
\begin{equation}
\begin{aligned}
P_s&=|\langle A_s|\psi\rangle|^2\\&=|\langle
A_s|A'_i\rangle|^2\left[1-\frac{A_w^ig_0t}{m\hbar}\langle
i|(\hat{p}^2-2m\hat{G})|i\rangle+o(g^2_0)\right]\,.
\end{aligned}
\end{equation}
Such an observable quality depends on the imaginary part of the weak
value, the atomic initial momentum uncertainty, and the
gravitational acceleration $\bar{g}$. Certainly, $P_s$ can be used
to test the effects of $g_0$.

Note that, the spatial motion of the selected atoms, i.e., the state
(22), also includes the information of parameter $g_0$. At a
glance, we formally write it as $|f\rangle\approx
\hat{V}_{t}\exp(ig_0tA_w\hat{H}_{\rm eff}/\hbar)|i\rangle$. Then,
the $x$-directional component reads
\begin{equation}
|f_{x}\rangle\approx e^{\frac{-it_r\hat{p}_x^2}{2m\hbar}}
e^{\frac{-t_i\hat{p}_x^2}{2m\hbar}}|\phi_x\rangle=
 \sum_{p_x}
e^{\frac{-it_rp_x^2}{2m\hbar}}e^{\frac{-t_ip_x^2}{2m\hbar}}\phi
(p_x)|p_x\rangle\,,
\end{equation}
with $t_r=t(1-g_0A^r_w)$ and $t_i=g_0tA_w^i$. Here, $A_w^r$ and
$A_w^i$ are respectively the real and imaginary parts of the weak value.
The initial state has been written as $|\phi_x\rangle=\sum_{p_x}\phi
(p_x)|p_x\rangle$ in the momentum Hilbert space, with $\phi(p_x)$
being the probability-amplitude of finding the momentum eigenstate
$|p_x\rangle$. The presentation (25) clearly shows that, the real
part of the weak value contributes phases to the atomic momentum
eigenstates, while the imaginary part changes the momentum
distribution of atoms. Whether using
Eq.~(22) or Eq.~(24) to measure $g_0$, the initially large momentum
uncertainty should be prepared. In the following section we
discuss the relevant issues.

\section{Implementing weak measurement based on the Kapitza-Dirac scattering of atoms}
\subsection{The state preparation}
First, we briefly review the atomic KD scattering. On the one hand,
it can generate the large momentum uncertainty of atoms (a necessary
condition for measuring the coupling $|e\rangle\langle
e|\hat{p}^2$). On the other hand, the KD scattering can be also
utilized to realize the desired postselection of the atomic internal
state. Note that, the KD scattering has been well studied in atom
optics~\cite{atomoptics}. It can be regarded as a position-dependent
ac Stark effect of atoms in the standing light wave. Here, we induce
an auxiliary atomic level, namely $|a\rangle$, see Fig.~1. Within
the near-resonant regime of transition
$|g\rangle\rightleftharpoons|a\rangle$, the ac Stark effect can be
described by a Hamiltonian of
$\hat{H}_{ac}=-\hbar\Omega\sin^2(kx)(|a\rangle\langle
a|-|g\rangle\langle g|)$. Where, $k$ is the wave number of the
applied light, $\Omega=\Omega^2_c/(\omega_l-\omega_a)$ describes the
frequency shift between the selected two levels. $\Omega_c$ is the
well-known resonant Rabi frequency, and $\omega_l-\omega_a$ is the
detuning between the light and the transition frequency of
$|g\rangle\rightleftharpoons|a\rangle$. Under the so-called
Raman-Nath approximation, the atomic wave function immediately after
the interaction is given by $|{\rm
KD}\rangle=\exp[-i\hat{p}^2\tau/(2m\hbar)]\exp(-i\hat{H}_{ac}\tau/\hbar)
|\psi_{0}\rangle$~\cite{KD1}. Here, $\tau$ and $|\psi_{0}\rangle$
are the effective interaction time and the initial state of the
system, respectively.

\begin{figure}[tbp]
\includegraphics[width=8.5cm]{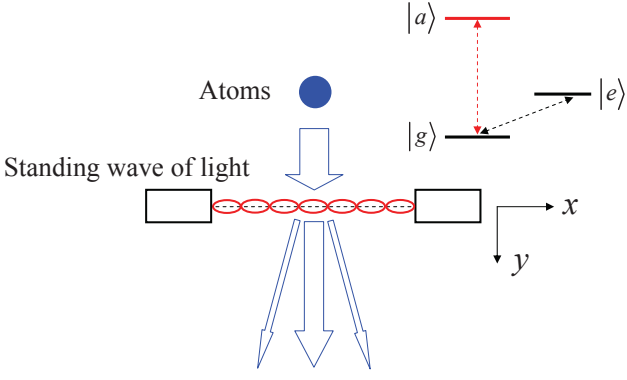}
\caption{State-selective scattering of the three-level V-type
atoms. An incident Gaussian beam of atoms is scattered transversely
by the standing wave of light, resulting in a large momentum
uncertainty of spatially coherent atoms. The atomic KD scattering
occurs within the near-resonant regime of optical transition
$|g\rangle\rightleftharpoons|a\rangle$. When atoms are initially in
the internal state $|e\rangle$, the scattering is negligible. This
provides an effective approach to measure the qubits $|g\rangle$ and
$|e\rangle$ by observing the external c.m. momenta of atoms.}
\end{figure}

Considering the atomic Gaussian beam incident, the external
$x$-directional motion is initially in the state of
\begin{equation}
|0\rangle =c\int_{-\infty}^\infty e^{\frac{-x^2}{4\Delta^2}}
|x\rangle dx=c'\int_{-\infty}^\infty e^{\frac{-p_x^2}{4\sigma^2}}
|p_x\rangle dp_x\,.
\end{equation}
Here, $c=(2\pi\Delta^2)^{-1/4}$ and $c'=(2\pi\sigma^2)^{-1/4}$ are
the normalized coefficients, and $|x\rangle$ and $|p_x\rangle$ are
respectively the atomic position and momentum eigenstates. The
quantities $\Delta$ and $\sigma=\hbar/(2\Delta)$ are the atomic
position and momentum uncertainty, respectively. Obviously, the
large position uncertainty corresponds to a small momentum
uncertainty. If the internal state of the atoms is initially in the ground
state $|g\rangle$, the scattered state reads~\cite{KD1}
\begin{equation}
|{\rm KD}\rangle=e^{-i\Omega \tau\sin^2(kx)}|0\rangle
=e^{i\eta\cos(2kx)}|0\rangle\,.
\end{equation}
In short, we have denoted $\Omega \tau/2=\eta$ and neglected the
phase $\exp[-i\hat{p}_x^2\tau/(2m\hbar)]\exp(-i\eta)$. Expanding
$\exp[i\eta\cos(2kx)]$ by the Bessel functions $J_n(\eta)$ of the
first kind, the atomic external motion reads
\begin{equation}
|{\rm KD}\rangle=\sum_{n=-\infty}^{\infty}\phi(n)
e^{i2nkx}|0\rangle\,,
\end{equation}
with $\phi(n)=i^nJ_n(\eta)$. Above, the term $\exp(i2nkx)$ acts as a
momentum-displacement operator in the $x$ direction, i.e.,
\begin{equation}
e^{i2nkx}|0\rangle=|n\rangle=
c'\int_{-\infty}^{\infty}e^{-\frac{(p_x-2n\hbar
k)^2}{4\sigma^2}}|p_x\rangle dp_x\,.
\end{equation}

If the initial momentum uncertainty is small, i.e., $\sigma\ll\hbar
k$, the states $\{|n\rangle\}$ can be regarded as the orthonormal
basis, i.e., $\langle n'|n\rangle=\delta_{n',n}$ (with
$\delta_{n',n}$ being the Kronecker delta function). Numerically,
considering the atomic position uncertainty $\Delta=1~\mu$m and the
wavelength $\lambda=0.4~\mu$m of a visible light, we have $\hbar
k/\sigma=4\pi\Delta/\lambda\approx31$ and $\langle
0|1\rangle=\exp[-(\hbar k)^2/(2\sigma^2)]=\exp(-480.5)$. It can be
further found that $\langle n'|\hat{p}_x^2|n\rangle \approx(2n\hbar
k)^2\delta_{n',n}$, such that the momentum uncertainty of the
scattered atoms is calculated as
$\langle\phi_x|\hat{p}_x^2|\phi_x\rangle=(2\hbar
k)^2\vartheta$~\cite{KD1,KD2,KD3}. Where, the quantity
$\vartheta=\sum_{-\infty}^{\infty}n^2J_n^2$ can be directly computed
by the numerical method, for example, $\vartheta=50$ with $\eta=10$.

The detection of momentum states $\{|n\rangle\}$ is similar to that
of the classical particles, by directly observing their
distinguishable paths. Considering the free diffraction, we replace
Eq.~(28) by
\begin{equation}
|\phi_x\rangle=e^{\frac{-it_0\hat{p}^2_x}{2m\hbar}}|{\rm
KD}\rangle=\sum_{n=-\infty}^{\infty}\phi(n)|\tilde{n}\rangle\,.
\end{equation}
Here, $t_0=\tau+\tau'$, with $\tau'$ being the time after the light
driving. For the time-dependent momentum state
$|\tilde{n}\rangle=\exp[-it_0\hat{p}_x^2/(2m\hbar)]|n\rangle$, its
wave packet $|\langle x|\tilde{n}\rangle|^2$ spreads with a quantity
$\Delta_d=\hbar t_0/(2m\Delta)$ which is smaller than the shift
$D_n=4n\pi\hbar t_0/(m\lambda)$ of the wave-packet center (because
$\Delta>\lambda$). Hence, the states $\{|\tilde{n}\rangle\}$ can be
distinguishable in position space when the time $\tau'$ is
sufficiently long, i.e., $D_n\gg\sqrt{\Delta^2+\Delta^2_d}$. The
atoms beam in momentum state $|\tilde{n}\rangle$ can be directly
observed by placing a detector at the position $D_n$ (with a
collecting region $\sim\sqrt{\Delta^2+\Delta^2_d}$).

\subsection{The postselection}
The desired postselection of internal state $|A_s\rangle$ can be
realized by the following two steps. First, we perform a
single-qubit rotation $\hat{R}=\exp[-i\theta(|e\rangle\langle
g|+|g\rangle\langle e|)]$ to the atomic internal state by the Raman
beams. The parameter $\theta$ is a controllable quantity which is
proportional to the power and duration of the drivings. For
simplicity, we assume that the single-qubit operations do not affect
the $x$-directional c.m. motions. This can be realized by properly
selecting the direction of laser irradiating, e.g., the $y$
direction. Second, we use another laser to realize the KD scattering
of atoms, i.e., the Hamiltonian $\hat{H}_{ac}$. Since the effective
interaction occurs between the transition
$|a\rangle\leftrightharpoons|g\rangle$, the scattering is
state-selective. If the atom is in state $|e\rangle$, there is no KD
scattering. However, the atoms in state $|g\rangle$ will be
scattered and consequently deviate from their original trajectory to
be detected. Worthy of note is that the present postselection is
similar to that in the original work~\cite{AAV} by Aharonov \emph{et
al}. There, a Stern-Gerlach device is arranged to implemented the
postselection of electronic spin. It couples the spin to the
orbital motions of the electron. Consequently, they select an orbital
motion to realize the postselection of the electronic spin state. Here,
we use a state-dependent KD scattering to implement the desired
postselection of the atomic internal state. Supposing the total
duration of the above two operations is sufficiently short (i.e.,
the interaction of $g_0$ in this stage is negligible), the desired
postselection operation $\langle g|\hat{R}$ can be realized. For
such a postselection, the weak value reads
\begin{equation}
A_w =\frac{\langle g|\hat{R}|e\rangle\langle e|A'_i\rangle}{\langle
g|\hat{R}|A'_i\rangle}=\frac{\beta}{\beta+ie^{i\omega
t}\alpha\cot(\theta)}
\end{equation}
with the initial state $|A_i\rangle=\alpha|g\rangle+\beta|e\rangle$.
Obviously, the real or imaginary part of the above weak value can be
nonzero. That is, the single-qubit operations in state preparation
and postselection will change the final results of detection.
In particular, if $\alpha=\beta=1/\sqrt{2}$, $\theta=3\pi/4$, and
$\omega t\rightarrow(2n-1/2)\pi$~\cite{Miao}, we have
$A_w^i\rightarrow\infty$. Of course, the relevant qualities such as
the durations of laser pulses should be exactly controlled in
experiments, like the well-known Ramsey
interferometers~\cite{atomoptics}.

\subsection{The weak-value amplification}
\begin{figure}[tbp]
\includegraphics[width=8.5cm]{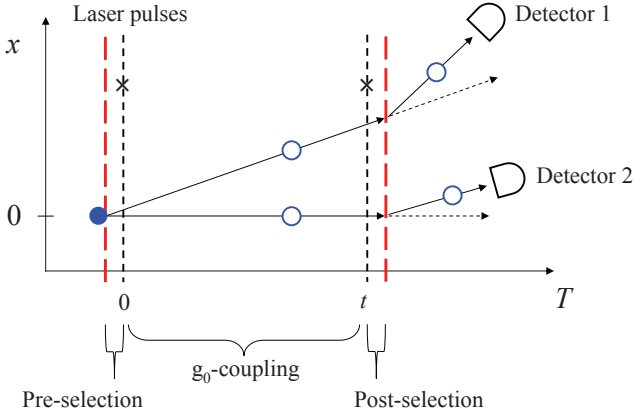}
\caption{A flow chart of weak measurement. The preselection (state
preparation) includes two steps, the KD scattering for atomic
external state preparation and the single-qubit operation for
internal superposition state generation. After a duration of
$g_0$ coupling, the postselection is implemented by the other
single-qubit operation and KD scattering. The momentum distribution
of the scattered atoms depends on the parameter $g_0$ and the
imaginary part of the weak value. By comparing the intensities of the
selected two atomic beams the parameter $g_0$ could be estimated.}
\end{figure}
The considered weak measurement process is diagrammatically shown in
Fig.~2. The atoms undergo the following several stages: (a)
standing light wave induced KD scattering in the $x$ direction, (b)
single-qubit operation performed by the $y$-directional laser beams,
(c) the $g_0$ couplings, (d) postselection implemented by the
other single-qubit operation and KD scattering, and (e) the final
detection. In fact, steps (a) and (b) are those for the initial
state preparations of both the atomic internal and external degrees of
freedom.

Regarding (30) as the external initial state of atoms, the
single-qubit operation was performed to generate the desired
internal state, i.e., $|g\rangle\rightarrow|A_i\rangle$. After a
duration of $g_0$ coupling, the postselection was performed on the
atomic internal state, resulting in state (22). Directly, putting
the prepared state (30) into Eq.~(22), we have
\begin{equation}
\begin{aligned}
|f\rangle=&\langle
A_s|A'_i\rangle\hat{V}_{t}\left[1+\frac{iA_wg_0t}{2m\hbar}
(\hat{p}^2-2m\hat{G})\right]\sum_{n=-\infty}^{\infty}\phi(n)
|\tilde{n}\rangle\\&\otimes |\phi_y\rangle|\phi_z\rangle\,.
\end{aligned}
\end{equation}
Under the approximation $\langle
\tilde{n}'|\hat{p}_x^2|\tilde{n}\rangle=\langle
n'|\hat{p}_x^2|n\rangle\approx(2n\hbar k)^2\delta_{n',n}$, the
probability of finding the momentum state $|\tilde{n}\rangle$ is
calculated as
\begin{equation}
\begin{aligned}
P_n&=P_s\frac{|\langle \tilde{n}|f\rangle|^2}{\langle
f|f\rangle}=|\langle \tilde{n}|f\rangle|^2 \\
&=|\langle
A_s|A'_i\rangle|^2J_n^2\left\{1-\frac{g_0tA_w^i}{m\hbar}\left[(2n\hbar
k)^2+\langle yz\rangle\right]\right\}
\end{aligned}
\end{equation}
Here, the postselection probability $P_s$ is induced, i.e., our
statistics is based on the entire atomic ensemble. The term $\langle
yz\rangle= \langle i|(\hat{p}_y^2+\hat{p}_z^2-2m\hat{G}) |i\rangle$
refers to $y$ and $z$ directional motions. Its contribution for
Eq.~(33) is small because momentum uncertainties in the $y$ and $z$
directions were not enhanced. Due to the limited interaction time
$t$, the contribution from gravity could be also negligible.
Therefore, Eq.~(33) can be reduced as
\begin{equation}
P_n=|\langle A_s|A'_i\rangle|^2J_n^2(1-g_0\omega_ktn^2A^i_w)\,,
\end{equation}
by neglecting the term $\langle yz\rangle$, where $\omega_k=4\hbar
k^2/m$. The above equation indicates that the imaginary part of the weak
value will change the momentum distribution of the post-selected
atoms. This provides us an alternative approach to measure the
parameter $g_0$. As we mentioned earlier, the postselection
probability $P_s$ is also an effective measurement for $g_0$, if the
number $N$ of the input atoms is precise. Similarly, by neglecting
the contribution from $y$- and $z$-directional motions, we have
\begin{equation}
P_s=|\langle A_s|A'_i\rangle|^2(1-g_0\omega_kt\vartheta A^i_w)\,,
\end{equation}
which satisfies the normalized condition $\sum_{n=-\infty}^{\infty}
P_n=P_s$.

The usefulness of the weak value is significant: it can amplify the $g_0$ signal with the respective of backgrounds [i.e., the first terms in
Eqs.~(34) and (35)]. Generally, the output of an atom-detector can
be described by the equation
$I_n=(N+\xi_s)P_n+(N+\xi_s)\xi_dP_n+\varepsilon$. Here, $\xi_s$ is
an uncertainty of input atoms (namely, the source-noise), $\xi_d$
denotes the uncertainty of detection efficiency (such as the
detector saturation). In other words, $\xi_d$ is the noise which is
proportional to the amount of received atoms. $\varepsilon$ denotes
the dark-counting which is $N+\xi_s$ independent. A single atom can
be easily observed by its resonance fluorescence, so that the
dark-counting $\varepsilon$ is negligible. The main noises in the
atomic system are $\xi_s$ and $\xi_d$. Using the weak-value
amplification, the background-induced noises, e.g., $N|\langle
A_s|A'_i\rangle|^2\xi_d$, will be significantly suppressed.

Comparing to postselection probability $P_s$, the momentum
distribution $P_n$ of the post-selected atoms is more useful for
suppressing the source-noise. One can use two detectors to measure
such a distribution, for example,
\begin{equation}
\begin{aligned}
\frac{I_n}{I_{0}}&=\frac{(1+\xi_{1d})P_n}{(1+\xi_{2d})P_{0}}
\\&\approx(1+\xi_{1d}-\xi_{2d})
\left(1-g_0\omega_ktn^2A^i_w\right)\frac{J_n^2}{J_{0}^2}\,.
\end{aligned}
\end{equation}
Where, the source-noise (and the term $|\langle A_s|A'_i\rangle|^2$)
has been completely eliminated, and
$(1+\xi_{1d})/(1+\xi_{2d})\approx1+\xi_{1d}-\xi_{2d}$ is due to the
noises of two detectors (denoted by, respectively, $\xi_{1d}$ and
$\xi_{1d}$). The imaginary part of the weak value can be an evidence for
the coupling of $|e\rangle\langle e|p^2$. If there is no such
coupling (i.e., replacing $|e\rangle\langle e|$ by the unit operator
$\hat{1}$), the weak value $A_w^i=0$ and consequently the Eq.~(36)
is $g_0$ independent. In principle, when
$g_0\omega_ktn^2A^i_w>\xi_{1d},\,\xi_{2d}$, the signal of $g_0$
should be detectable.

Both Eq.~(34) and (35) indicate $g_0\omega_kt$ is a very crucial
quantity for measuring the parameter $g_0$. Such a quality depends
on $k^2$, so that using the high-frequency light scattering can
significantly enhance the desired signal of $g_0$. Of course,
$g_0t=\hbar\omega t/(mc^2)$ is the most basic quality. Where, the
interaction duration $t$ is limited within the coherent times of the
atomic qubit. Hence, the optical atomic clocks are perhaps the good
candidates for the relatively large $g_0t$ gain, because they have
the qualities of both large transition frequency $\omega$ and long
lifetime $t$~\cite{Ca1,Ca2,Ca3}. Considering the usually used
transition $^{1}S_0-^{3}P_1$ of a calcium clock, the resonant
frequency and the lifetime of the excited state $^{3}P_1$ are
respectively $\omega\approx4.6\times10^{14}$~Hz and
$t\approx0.4$~ms~\cite{Ca1}, such that
$g_0t\approx3.2\times10^{-15}$~s. The relatively strong KD
scattering occurs at the transition $^{1}S_0-^{1}P_1$ of calcium,
and where the wavelength of the driving light is on the order of
$\lambda=2\pi/k\approx0.4~\mu$m. Using these parameters, we have
$\omega_k\approx1.6$~MHz and $g_0\omega_kt\approx5\times10^{-9}$.
Considering a KD scattering with $\eta=10$, the probability for
finding momentum $|n=10\rangle$ is $J^2_{10}(\eta)\approx0.04$. As a
consequence, $g_0\omega_ktn^2A_w^i\approx5\times10^{-3}$ is obtained
by selecting a large weak value $A_w^i=1\times10^4$. Such a signal may be
detectable if the noises $\xi_{1d}$ and $\xi_{2d}$ are smaller than
$5\times10^{-3}$. Note that, the present weak measurement is also
suitable for some other scattering mechanisms, for example,
preparing the atomic initial state with more large momentum
uncertainty by the multi-step scattering of lights. In principle,
this will generate more significant effects of $g_0$ than the present
one.

\section{Conclusion}
In this theoretical work, we proposed a weak value scheme to measure
the effect of mass-energy equivalence in the system of coherent atoms. First, we
reviewed briefly the principle of quantum weak measurement and its
usefulness in the weak signal detections. Second, we presented an
approach to derive the dynamic evolution of an atom whose internal
and external degrees of freedom are coupled due to the mass-energy equivalence. After a
duration of such a coupling, the weak value is obtained by
post-selecting an atomic internal state. It is shown that the
imaginary part of the weak value can change the momentum distribution
of the selected atoms. This allowed us to measure the coupling of mass-energy equivalence in
the momentum space of atoms. Third, we used the KD scattering to
prepare and post-select the atoms (with the help of two single-qubit
rotations). Preparing atomic momentum states by the light-induced
scattering is optimal for measuring the coupling of
$|e\rangle\langle e|\hat{p}^2$, because the scattering can generate
a relatively large momentum uncertainty of coherent atoms. Certainly,
observing the effect of this coupling is still difficult by using
even advanced optical atomic clocks. Under this situation, the
weak value would be useful. It can significantly amplify the signals
above the background noises, making the detection of weak signals
feasible. Finally, we hope the present studies will
encourage further studies on the detection of weak signals.

{\bf Acknowledgements}: This work was partly supported by the National
Natural Science Foundation of China, Grants No. 11204249 and No.
11547311.
\\

\end{document}